\documentclass[a4paper]{article}

\usepackage{INTERSPEECH2022}
\usepackage{amsmath,comment,tablefootnote,url,cite}

\DeclareMathOperator{\Aux}{NN^{Aux}}
\DeclareMathOperator{\Ext}{NN^{Ext}}
\def\y{{\mathbf y}}
\def\x{{\mathbf x}}
\def\xs{{\mathbf{x}^s}}

\def\xhat{\hat{\mathbf x}^{s}}

\def\a{\mathbf{a}^s}

\def\n{{\mathbf n}}
\def\e{{\mathbf e}}
\def\es{\mathbf{e}^s}

\usepackage[acronym,shortcuts]{glossaries}
\glsdisablehyper
\newacronym{SDR}{SDR}{signal-to-distortion ratio}
\newacronym{SDRi}{SDRi}{signal-to-distortion ratio improvement}
\newacronym{SNR}{SNR}{signal-to-noise ratio}
\newacronym{SI-SDR}{SI-SDR}{scale-invariant \gls{SDR}}
\newacronym{SI-SNR}{SI-SNR}{scale-invariant \gls{SNR}}
\newacronym{NN}{NN}{neural network}
\newacronym{NNs}{NNs}{\gls{NN}s}
\newacronym{mAP}{mAP}{mean average precision}
\newacronym{GAN}{GAN}{generative adversarial network}
\newacronym{PIT}{PIT}{permutation invariant training}
\newacronym{IS}{IS}{inactive speaker}
\newacronym{AS}{AS}{active speaker}
\newacronym{FA}{FA}{False alarm}
\newacronym{MD}{MD}{Miss detection}
\newacronym{EER}{EER}{Equal error rate}

\newacronym{DNN}{DNN}{deep \gls{NN}}

\newacronym{TSE}{TSE}{Target speech extraction}
\newacronym{TSEIS}{TSE-IS}{\gls{TSE} with internal \gls{IS} detection}
\newacronym{TSEV}{TSE-V}{\gls{TSE}+Verification}
\newacronym{ConvTasNet}{Conv-TasNet}{fully-convolutional time-domain audio separation network}
\newacronym{STFT}{STFT}{short-time Fourier transform}
\newacronym{iSTFT}{iSTFT}{inverse \gls{STFT}}
\newacronym{FiLM}{FiLM}{Feature-wise Linear Modulation}
\newacronym{ROC}{ROC}{receiver operating characteristic}
\newacronym{DET}{DET}{detection error tradeoff}
\newacronym{AUC}{AUC}{area under the curve}
\newacronym{PANN}{PANN}{pretrained audio \gls{NN}}
\newacronym{FSD}{FSD}{free sound dataset}

\title{Listen only to me! How well can target speech extraction handle false alarms?}
\name{Marc~Delcroix$^1$, Keisuke Kinoshita$^1$, Tsubasa Ochiai$^1$, Katerina Zmolikova$^{2}$, Hiroshi Sato$^1$, Tomohiro Nakatani$^1$}
\address{
$^1$NTT Corporation, Japan, $^2$Brno University of Technology, Speech@FIT, Czechia\thanks{Katerina Zmolikova was partly supported by European Union's Horizon 2020 project No. 833635 ROXANNE.}}
\email{marc.delcroix@ieee.org}

\begin{document}
\setlength{\belowdisplayskip}{4pt} \setlength{\belowdisplayshortskip}{4pt}
\setlength{\abovedisplayskip}{4pt} \setlength{\abovedisplayshortskip}{4pt}

\maketitle
\begin{abstract}
Target speech extraction (TSE) extracts the speech of a target speaker in a mixture given auxiliary clues characterizing the speaker, such as an enrollment utterance. TSE addresses thus the challenging problem of simultaneously performing separation and speaker identification. There has been much progress in extraction performance following the recent development of neural networks for speech enhancement and separation. Most studies have focused on processing mixtures where the target speaker is actively speaking. However, the target speaker is sometimes silent in practice, i.e., inactive speaker (IS). A typical TSE system will tend to output a signal in IS cases, causing false alarms. It is a severe problem for the practical deployment of TSE systems. This paper aims at understanding better how well TSE systems can handle IS cases. We consider two approaches to deal with IS, (1) training a system to directly output zero signals or (2) detecting IS with an extra speaker verification module. We perform an extensive experimental comparison of these schemes in terms of extraction performance and IS detection using the LibriMix dataset and reveal their pros and cons.  
\end{abstract}

\section{Introduction}
Enhancing a speech signal corrupted by interfering speakers has been one of the major challenges of speech signal processing.
One way to tackle this problem is to use speech separation~\cite{makino2007blind}, which separates a speech mixture into all its sources. Research in speech separation has progressed rapidly with the advent of deep learning~\cite{hershey2016deep,yu2016permutation,luo2018tasnet}. However, there are two fundamental limitations with most separation techniques. First, separation requires knowing or estimating the number of sources in the mixture. Then, there is a global permutation ambiguity; the mapping between outputs speakers is arbitrary.

\gls{TSE}~\cite{zmolikova2019Journal} is an alternative to enhance speech in a mixture. \gls{TSE} focuses on extracting only a target speaker's speech instead of separating all sources by exploiting a speaker clue to identify that speaker~\cite{zmolikova2017speaker,delcroix2018single,Wang_voicefilter19,zmolikova2019Journal,xu2019time,delcroix2020improving,ephrat2018looking,afouras2018conversation,Chen2018DeepEN,xiao2019single,Jansky_20,Gu2019}. For example, we can use an enrollment utterance consisting of a short recording containing only the voice of the target speaker~\cite{zmolikova2017speaker,delcroix2018single,Wang_voicefilter19}. Because \gls{TSE} estimates only the speech of the target speaker, it naturally alleviates the issues of separation systems, i.e., the processing is independent of the number of sources in the mixtures, and there is no speaker ambiguity at the output.

We can realize \gls{TSE} using a \gls{NN} conditioned on the target speaker clue, which directly estimates the target speech from the mixture~\cite{zmolikova2017speaker,delcroix2018single,Wang_voicefilter19,zmolikova2019Journal,xu2019time,delcroix2020improving}. Such a \gls{TSE} system must perform thus both \emph{separation} and \emph{speaker identification} internally. 
Most studies about \gls{TSE} have assumed the target speaker was always actively speaking in the mixtures, i.e., \emph{\gls{AS}} case. However, we argue that measuring \gls{TSE} performance in such conditions does not fully represent the \emph{speaker identification} capabilities of \gls{TSE} systems. Indeed, in practice, a target speaker may be silent, i.e., \emph{\gls{IS}} case. In such a case, a \gls{TSE}  should output nothing or a zero signal. However, a \gls{TSE} system trained only on \gls{AS} conditions would always try to output a speech-like signal, which would cause \emph{false alarms} or false positive. It is thus essential to consider \gls{IS} conditions in the design and evaluation of \gls{TSE} systems.

There have been only a few works dealing with the \gls{IS} issue of \gls{TSE}~\cite{zhang20m_interspeech,borsdorf21_interspeech,Zhang2021}. These works offer two different strategies to address the problem. The \gls{TSEIS} scheme trains a \gls{TSE} system to directly output zero signals for \gls{IS} cases by including \gls{IS} samples during training~\cite{zhang20m_interspeech,borsdorf21_interspeech}. The \gls{TSEV} scheme combines \gls{TSE} with speaker verification and detects \gls{IS} samples when the extracted signals do not match the target speaker characteristics of the enrollment~\cite{Zhang2021}\footnote{Note that in \cite{Zhang2021} the goal is not \gls{TSE} but speaker verification.}.  \gls{TSEIS} is a simpler system than \gls{TSEV}, but it is potentially easier to control false alarm and \emph{miss detection}\footnote{Here miss detection or false negative means that the \gls{TSE} systems wrongly predicted an \gls{AS} as \gls{IS}.} with \gls{TSEV}. However, these schemes have not been compared, and their impact on \gls{TSE} performance has not been fully revealed.
In this paper, we address this shortcoming and perform a comprehensive comparison in terms of the detection of \gls{IS} and extraction performance in order to answer the following question: \emph{How well can \gls{TSE} systems handle \gls{IS} samples?}

The contribution of this paper is as follows: 
(1) We propose two simple implementations of the \gls{TSEIS} and \gls{TSEV} schemes based on the SpeakerBeam \gls{TSE} framework~\cite{delcroix2020improving}, and perform an comprehensive experimental comparison in terms of extraction and \gls{AS}/\gls{IS} detection performance. 
(2)  We reveal that a \gls{TSEIS} system trained with a modified \gls{SNR} loss can predict \gls{IS} in about 90~\% of the cases but also significantly increases the number of extraction failures for \gls{AS} cases. 
(3) We show that we can build a \gls{TSEV} system from a \gls{TSE} system trained only with \gls{AS} samples. Such a \gls{TSEV} system can  detect \gls{AS}/\gls{IS} better than a \gls{TSEIS}, while maintaining high extraction performance. (4) Finally, we reveal that the enrollment duration impacts moderately extraction performance but greatly affects \gls{AS}/\gls{IS} detection errors of \gls{TSEV}. With enrollment of 15 sec or more, we can achieve \gls{AS}/\gls{IS} detection with a \gls{EER} of about 5~\%.
These results demonstrate the potential of current \gls{TSE} systems to detect and extract a target speaker.

\section{Related works}
Prior works~\cite{zhang20m_interspeech,borsdorf21_interspeech} considered \gls{TSE} with \gls{IS} cases. They introduced a modified \gls{SI-SNR}~\cite{roux2019sdr} loss to allow training a \gls{TSEIS} system with \gls{IS} samples. However, using a scale-invariant loss makes the output scale arbitrary. It may thus be challenging to determine if the output can actually be considered an \gls{AS} or not. Besides, it is unclear how well the system proposed in~\cite{borsdorf21_interspeech} can detect \gls{IS} cases since the approach was only evaluated with signal extraction measures and not in terms of \gls{AS}/\gls{IS} detection. In contrast, we propose using a modified \gls{SNR} loss to train our \gls{TSEIS} system, which preserves the scale at the output of the system and allows thus performing the \gls{AS}/\gls{IS} detection based on the attenuation from the mixture. 

There have been two prior studies combining \gls{TSE} with speaker verification~\cite{rao19_interspeech, Zhang2021}, which are related to \gls{TSEV}. Both works aimed at improving speaker verification for speech in a mixture and used a \gls{TSE} system as a pre-processing. 
However, as their goal was speaker verification and not \gls{TSE}, they did not evaluate their systems in terms of extraction performance although, e.g., miss detection errors caused by zeroing out the output when \gls{AS} cases are detected as \gls{IS} can have a severe impact on extraction performance.

\section{TSE problem and baseline system}
\label{sec:baseline}
\subsection{Problem formulation}
\gls{TSE} aims at extracting speech of a target speaker, $\xs \in \mathcal{R}^T$ from a mixture $\y \in \mathcal{R}^T$ defined as,
\begin{align}
    \y = \xs + \sum_{i \neq s} \x^i + \n,
\end{align}
where $\x^i$ and $\n$ represent the interference speech and the background noise signals, respectively. $T$ is the duration of the signal. We assume having an enrollment utterance of the target speaker, $\a \in \mathcal{R}^{T^a}$, of duration $T^a$.
Note that when the target speaker is active, $\xs$ is a speech signal, while when it is inactive, $\xs = \mathbf{0}$, where $\mathbf{0}$ denotes a vector of all zeros.

\subsection{SpeakerBeam}
\begin{figure}[tb]
\centering
  \centerline{\includegraphics[width=0.95\linewidth]{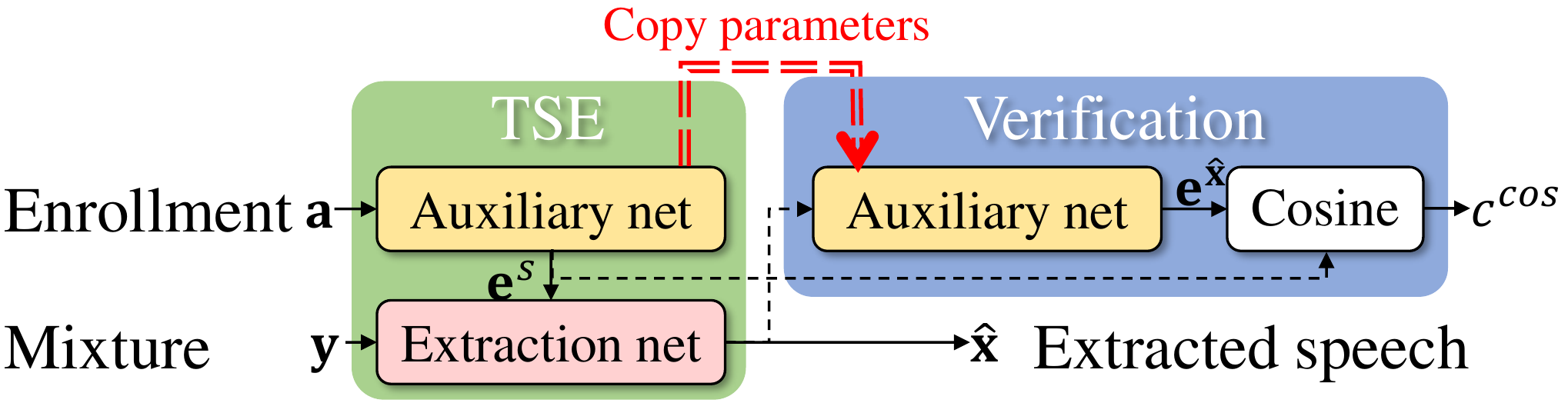}}
  \vspace{-2mm}
\caption{Overview of a \gls{TSE} system and its extension to \gls{TSEV}.}
\vspace{-4mm}
\label{fig:sys_overview}
\end{figure}
 
 We use time-domain SpeakerBeam~\cite{delcroix2020improving} as a basis for our study as it represents a typical enrollment-based neural \gls{TSE} system~\cite{Wang_voicefilter19,xu2019time}. 
 The left part of Fig.~\ref{fig:sys_overview} shows a diagram of the system. It consists of two modules. (1) An auxiliary \gls{NN} that computes a target speaker embedding, $\es \in \mathcal{R}^N$, from the enrollment, $\a$. (2) An extraction \gls{NN} that estimates the target speech from the mixture given the speaker embedding. The operation of the \gls{NN} is summarized as follows,
\begin{align}
    \es = & \Aux(\a),\\
    \xhat = & \Ext(\y, \es),
\end{align}
where $\xhat \in \mathcal{R}^T$ is the estimated target speech,  $\Aux(\cdot)$ and $\Ext(\cdot)$ are the auxiliary and extraction \glspl{NN}, respectively.

Both $\text{NN}^{\text{Aux}}(\cdot)$ and $\text{NN}^{\text{Ext}}(\cdot)$ consists of 1-D convolutional blocks as in the \gls{ConvTasNet}~\cite{luo2019conv}. The extraction \gls{NN} uses an element-wise multiplication~\cite{samarakoon2016subspace,delcroixIcassp19} to combine the embedding vector with the hidden representation obtained after the first convolutional block of the extraction \gls{NN}.

\subsection{Training objective for active speaker cases}
We train the auxiliary and extraction \glspl{NN} jointly, enabling learning speaker embeddings optimal for \gls{TSE}. Speech separation and \gls{TSE} systems are usually trained using a time-domain criterion such as \gls{SNR} or \gls{SI-SNR}~\cite{roux2019sdr,luo2019conv}. 
We chose to use a scale-dependent loss, to ensure that the system preserves the scale of the signals as it may be important to detect \gls{AS}/\gls{IS} samples. We use the negative thresholded \gls{SNR}~\cite{wisdom2020unsupervised} loss,
\begin{align}
\mathcal{L}^{\text{active}} (\xhat, \x^s )& =  - 10 \log_{10} \left( \frac{\| \x^s \|^{2}}{\| \x^s - \xhat \|^{2} + \tau \| \x^s \|^2} \right), \label{eq:ext_loss} 
\end{align}
where $\tau$ is a threshold that we set at $\tau = 10^{-3}$. It avoids that the low distortion training samples dominate the gradient. 
We train our baseline \gls{TSE} model with only \gls{AS} training samples.

\section{Handling inactive speakers}

\subsection{TSE-IS: Learning direct \gls{IS} detection with inactive loss}
\label{ssec:SpeakerBeam-IS}
The first approach for handling \gls{IS}, \gls{TSEIS}, consists of training a system to output zero signals for \gls{IS} cases. 
The loss functions derived from the \gls{SNR} such as Eq.~\eqref{eq:ext_loss} are ill-defined when the reference signal is zero. Thus, we cannot use it directly with \gls{IS} samples. This problem was first revealed for the training of separation systems that can accommodate a varying number of sources in a mixture~\cite{FUSS_wisdom2021s}, i.e., the number of sources can be less than the number of outputs of the separation system. In this case, a separation system needs to be able to output zero signals, which is similar to the \gls{IS} problem of \gls{TSE}. 

We propose to use a modified \gls{SNR} loss~\cite{FUSS_wisdom2021s},
\begin{align}
    \mathcal{L} (\xhat, \x^s, \y )= &\left\{ \begin{array}{lc}
        \mathcal{L}^{\text{active}} (\xhat, \x^s ), &  \text{if } \x^s \neq \mathbf{0},\\
        \mathcal{L}^{\text{inactive}} (\xhat, \y ), & \text{if } \x^s = \mathbf{0},
    \end{array} \right.
    \label{eq:active&inactive_loss}
\end{align}
where the inactive loss is given by, 
\begin{align}
    \mathcal{L}^{\text{inactive}} (\xhat, \y )=10 \log_{10} \left( \|  \xhat \|^{2} + \tau^{\text{inactive}} \| \y \|^2  \right),
\end{align}
and $\tau^{\text{inactive}}$ is a soft threshold set at $\tau^{\text{inactive} } = 10^{-2}$.
$\mathcal{L}^{\text{inactive}}$ consists of the denominator term of Eq. (\ref{eq:ext_loss}) with a different setting for the soft threshold (i.e., $\x^s$ replaced by $\y$). This loss minimizes the norm of the estimated signals, $\xhat$, for \gls{IS} cases, which makes them tend to zero.

We opt here for a scale-dependent \gls{SNR} loss, unlike \cite{borsdorf21_interspeech}, because we believe that the scale of the output signal may matter in practical applications to detect \gls{IS} cases. For example, we can evaluate how well the system could internally detect \gls{AS}/\gls{IS} cases by looking at the attenuation from the mixtures, $\mathcal{A}^{\text{mixture}}= 10 \log_{10} \left(\frac{\| \xhat \|^{2}}{\| \y \|^{2}} \right)$.
We can thus define a \gls{AS}/\gls{IS} classifier as,
\begin{align}
  c^{\text{Att}} = &\left\{ \begin{array}{lc}
       1, &  \text{if }  \mathcal{A}^{\text{mixture}} > \eta^{\text{Att}},\\
       0, &  \text{if }  \mathcal{A}^{\text{mixture}} \le \eta^{\text{Att}},
    \end{array} \right.
    \label{eq:classifier_att}  
\end{align}
where $\eta^{\text{Att}}$ is a threshold. The target speaker is considered active when $c^{\text{Att}}=1 $ and inactive when $c^{\text{Att}}=0 $. 

We introduced the above classifier to measure the \gls{AS}/\gls{IS} detection capability of the system, but in practice, we do not need it as \gls{TSEIS} performs the \gls{AS}/\gls{IS} detection internally and directly output a speech signal or a zero signal. 
There is thus no increase in computational complexity compared to an existing \gls{TSE} system. However, it allows little control to, e.g., balance the false alarms or miss detection errors of the system at test time. Besides, adding \gls{IS} cases during training may hurt the extraction performance for the \gls{AS} cases.

\subsection{TSE-V: Post \gls{AS}/\gls{IS} detection with speaker verification}
\label{ssec:SpeakerBeamV}
Another approach to handle \gls{IS} cases, shown in Fig. \ref{fig:sys_overview}, consists of using a \gls{TSE} system trained on \gls{AS} cases, which always tries to output a speech-like signal, and then perform post verification to check that the speech characteristics of the extracted speech, $\xs$, correspond to those of the enrollment. 

In this work, we propose using the auxiliary \gls{NN} of SpeakerBeam to compute a speaker embedding for the extracted speech, $\e^{\xhat} = \Aux(\xhat)$, since we showed in prior works that it could extract discriminative speaker embeddings\cite{zmolikova2019Journal}. 
We then make the  \gls{AS}/\gls{IS} decision by looking at the cosine similarity between the embeddings computed from the enrollment and from the extracted speech as,
\begin{align}
  c^{\text{Cos}} = &\left\{ \begin{array}{lc}
       1, &  \text{if }  \mathcal{C}(\e^{\xhat}, \es) > \eta^{\text{Cos}},\\
       0, &  \text{if }  \mathcal{C}(\e^{\xhat}, \es) \le \eta^{\text{Cos}},
    \end{array} \right.
    \label{eq:classifier_cos}  
\end{align}
where $\mathcal{C}(\cdot)$ is the cosine similarity, and $\eta^{\text{Cos}}$ is a threshold. We define an extracted signal after detection as $\bar{\x}^s =c^{\text{Cos}} \xhat$, which zeros out the samples detected as \gls{IS}.

Note that this approach checks whether the extracted speech matches the enrollment characteristics. It can thus possibly detect not only \gls{IS} but also extraction failures. Such failures occur when the \gls{TSE} system wrongly outputs the mixture or the interference speakers instead of the target speech.

Compared to \gls{TSEIS}, \gls{TSEV} increases the computational complexity slightly as it requires an additional pass through the auxiliary \gls{NN}. However, since the \gls{AS}/\gls{IS} detection is performed independently of the \gls{TSE} process, it allows better control at test time and also does not require the training of the \gls{TSE} module with \gls{IS} samples. Note that in contrast to our proposed \gls{TSEV}, the system proposed in~\cite{Zhang2021} used a pre-trained speaker embedding extractor and retrained it on extracted speech. We can view our \gls{TSEV} system as a simplified version of~\cite{Zhang2021}.

\section{Experiments}
We performed experiments using the LibriMix dataset~\cite{cosentino2020librimix}, which consists of noisy two-speaker mixtures derived from the LibriSpeech dataset~\cite{panayotov2015librispeech}. We used the open implementation of SpeakerBeam~\cite{spkbeam_code} based on the asteroid toolkit~\cite{Pariente2020Asteroid}.

\subsection{Dataset}
\begin{table}[tb]
    \centering
    \caption{Description of the dataset}
    \vspace{-3mm}
    \begin{tabular}{l@{}cccc@{}}
    \toprule
                        & Train-100k & Train-360k & Val & Test \\
    \midrule
    Nb. of mixtures     & 13900 & 50800 & 3000 & 3000 \\
    Nb. of Speakers     & 251   & 921   & 40    & 40 \\
    \bottomrule
    \end{tabular}
    \label{tab:dataset}
    \vspace{-3mm}
\end{table}

We performed experiments using the full-overlap (i.e., min version) two-speaker noisy mixtures of the LibriMix dataset.
Table \ref{tab:dataset} provides more details about the dataset. For each mixture, we randomly sampled enrollment utterances from the speakers in the mixture for \gls{AS} cases and from a different speaker for the \gls{IS} cases. 
In both cases, the enrollment differed from the utterances used in the mixture. 
At test time, we considered enrollment utterances from three speakers for each test mixture, i.e., two from the \glspl{AS} in the mixture and one from another speaker, i.e., \gls{IS}.

\subsection{Experimental settings}
We used the same \gls{NN} architecture for all experiments, which consists of the SpeakerBeam system provided in~\cite{spkbeam_code}, except that we used the training loss of Eq.~\eqref{eq:active&inactive_loss}. We followed a similar configuration as \gls{ConvTasNet}~\cite{luo2019conv}. We used blocks of eight stacked 1-D convolution layers for the auxiliary and extraction \glspl{NN}, repeated three times for the extraction \gls{NN}. We used an element-wise multiplication to combine the embedding vector with the hidden representation at the output of the first convolution block. We trained the systems for 200 epochs with the Adam optimizer~\cite{kingma2015adam}.

We compared the following four \gls{TSE} systems. 
\textbf{Baseline \gls{TSE}} corresponds to the baseline  system of Section \ref{sec:baseline}, which was trained with the train-100k set with only \gls{AS} samples. It does not perform neither internal nor post \gls{AS}/\gls{IS} detection. 
\textbf{\gls{TSEIS}} corresponds to the system described in Section \ref{ssec:SpeakerBeam-IS}, which was trained with the train-100k training set including 10~\% of \gls{IS} cases, i.e., we used an enrollment from a speaker not present in the mixture and a zero signal as the target for 10~\% of the training samples. 
\textbf{\gls{TSEV}} corresponds to the system described in Section \ref{ssec:SpeakerBeamV}. The \gls{TSE} module corresponds to the above baseline \gls{TSE} system trained with only \gls{AS} samples. At test time, we re-used the auxiliary \gls{NN} to compute the embedding vector for the extracted speech and performed \gls{AS}/\gls{IS} detection with Eq.~\eqref{eq:classifier_cos}. 
\textbf{\gls{TSEV}(360)} consists of the \gls{TSE} module of the above \gls{TSEV} system retrained on \gls{AS} samples of the train-360k dataset for 100 epochs. We use it to measure the impact of having a larger training set with more speakers. 

\subsection{Evaluation metrics}
We evaluated the systems in terms of the following evaluation metrics:
(1) \textbf{\gls{EER}} measures the \gls{AS}/\gls{IS} detection errors using the \gls{DET} curves shown in Fig. \ref{fig:DET} obtained with the classifiers of Eq.~\eqref{eq:classifier_att} and Eq.~\eqref{eq:classifier_cos} for \gls{TSEIS}  and \gls{TSEV}, respectively. 
(2) \textbf{\Gls{SDRi}} measures the extraction performance for the \gls{AS} cases using the BSS eval toolkit~\cite{vincent2006performance}. We report the \gls{SDRi} before and after \gls{AS}/\gls{IS} detection, i.e.,  using $\xhat$ or $\bar{\x}^s =c \xhat$, respectively, where $c$ is given by either Eq.~\eqref{eq:classifier_att} or Eq.~\eqref{eq:classifier_cos} using the threshold that gives the \gls{EER}. We do not need to compute $\bar{\x}^s$ for \gls{TSEIS}, but we perform it anyway to provide a fair comparison with \gls{TSEV}. \gls{SDRi}-after accounts for the impact of miss detection errors on the extraction performance. Note that samples detected as \gls{IS}  are replaced by a zero signal, thus resulting in a SDR of 0~dB. 
(3) \textbf{Failure rate (Fail)} is defined as $Fail = \frac{\mathit{NF}^{\text{AS}}}{N^{\text{AS}}}$, where $\mathit{NF}^{\text{AS}}$ is number of \gls{AS} samples with \gls{SDRi} below 1~dB and $N^{AS}$ is the total number of \gls{AS} samples. Failures happen when, e.g., the \gls{TSE} system extracts the wrong speaker, output the mixture or a zero signal when using \gls{TSEIS}.
(4) \textbf{Failure and miss detection rate (Fail\&Miss)} is defined as $Fail\&Miss = \frac{\mathit{NFM}^{\text{AS}}}{N^{\text{AS}}}$, where $\mathit{NFM}^{\text{AS}}$ is the number of \gls{AS} samples that result in extraction or detection errors, i.e., \gls{SDRi} below 1~dB, miss detection or both. It measures the total error rate for the \gls{AS} cases. For example, even if a sample is correctly detected as \gls{AS} its extraction performance may be low, and it should thus be considered as an error.
(5) \textbf{Attenuation (Att.)} measures the attenuation from the mixture, $\mathcal{A}^{\text{mixture}}$, defined in Section \ref{ssec:SpeakerBeam-IS}. It shows how well a \gls{TSE} system can output zero signals for \gls{IS} cases.

Note that we use only \gls{AS} samples to compute \gls{SDRi}, Fail and Fail\&Miss, but both \gls{AS} and \gls{IS}  for \gls{EER} and Attenuation.

\subsection{Experimental results}
\begin{table}[tb]
  \caption{Extraction and  detection performance with enrollment of average duration of 10~sec. The input SDR is -1.8~dB. }
  \vspace{-2mm}
  \label{tab:results}
  \centering
  \begin{tabular}{@{}l@{}cccc@{}}
    \toprule
&  \gls{SDRi} before(after) & Fail&  \gls{EER}&  Fail\&Miss \\

& detection [dB] $\uparrow$  & [\%] $\downarrow$& [\%] $\downarrow$ &[\%] $\downarrow$ \\
  \midrule
Baseline TSE &  12.4  ( na )&  3.4& -&- \\
TSE-IS&  10.8 (11.4)&  8.6&  11.6&  13.4\\
TSE-V&  12.4  (11.9)&  3.4&  8.9&  10.5\\
\midrule
TSE-V(360)&  13.6 (13.1)&  1.7&  6.3&  7.1\\

\bottomrule
  \end{tabular}
  \vspace{-2mm}
\end{table}

\begin{figure}[tb]
\centering
  \centerline{\includegraphics[width=0.95\linewidth]{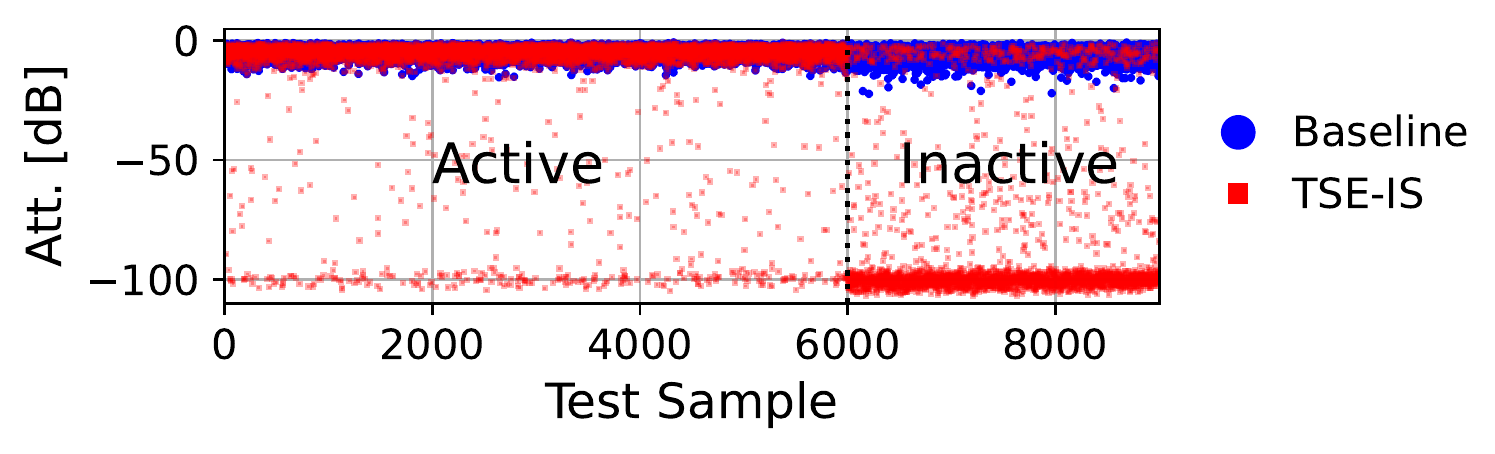}}
\vspace{-3mm}
\caption{Attenuation for each test sample. The first 6000 samples correspond to \gls{AS} and the last 3000 to \gls{IS} samples.}
\vspace{-4mm}
\label{fig:attenuation}
\end{figure}

Table \ref{tab:results} shows the extraction and \gls{AS}/\gls{IS} detection results for the different systems using enrollment of average duration of 10 sec. Figure \ref{fig:attenuation} shows the attenuation with respect to the mixture, $\mathcal{A}^{mixture}$, as a function of the samples in the test set. 

The baseline \gls{TSE} system, which was trained only with \gls{AS} samples, achieves 12.4~dB \gls{SDRi} 
and only 3.4~\% of failures. However, as seen in Fig. \ref{fig:attenuation}, the attenuation values remain in a similar range for \gls{AS} and \gls{IS} samples, meaning that it always outputs some signal even for \gls{IS} cases, causing many false alarms.

\begin{figure}[tb]
\centering
  \centerline{\includegraphics[width=0.95\linewidth]{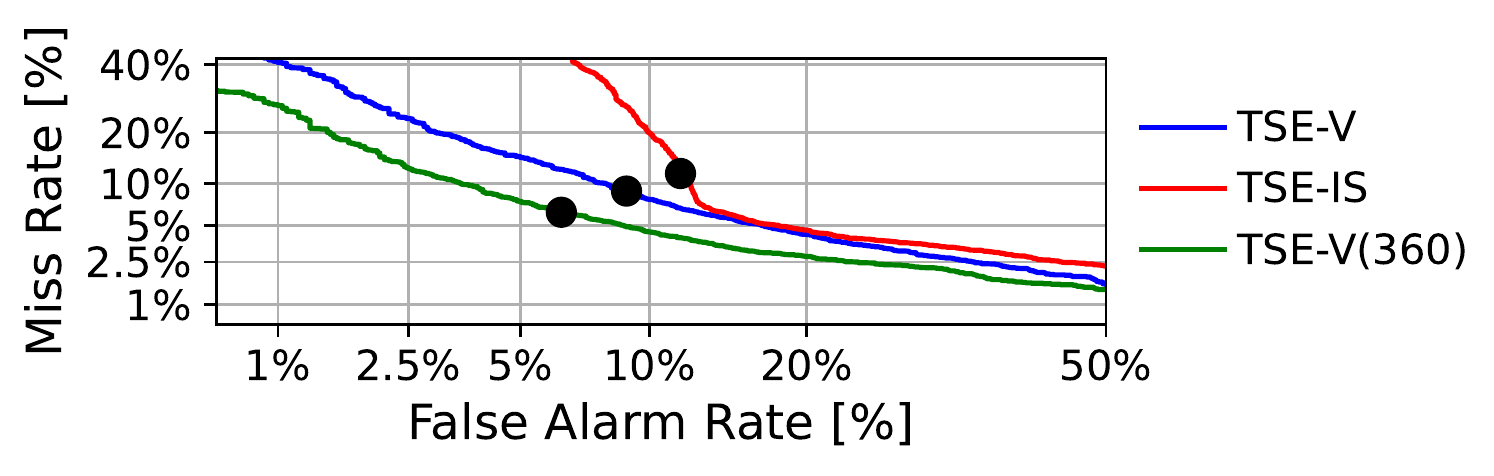}}
  \vspace{-3mm}
\caption{\gls{DET} curves for \gls{AS}/\gls{IS} detection with \gls{TSEV} and \gls{TSEIS}. The black circles indicate the \gls{EER}. }
\vspace{-3mm}
\label{fig:DET}
\end{figure}

The \gls{TSEIS} system, which we trained with \gls{IS} samples, can output zero signals. We observe in Fig.\ref{fig:attenuation} that the attenuation is around -100~dB for most \gls{IS} cases while it remains close to 0~dB for most \gls{AS} cases. This confirms that \gls{TSEIS} can internally perform \gls{AS}/\gls{IS} detection. However, we also observe that learning with \gls{IS} impacts extraction performance for \gls{AS} cases. Indeed, around 10~\% of the \gls{AS} test samples have attenuation around -100~dB (i.e., miss detection). Consequently, the failure rate is high, i.e., close to 9~\%, and the average \gls{SDRi} lower than the baseline. The impact on \gls{SDRi} may be exaggerated as it includes miss detection errors, i.e., samples where the system wrongly outputs a signal close to zero. The \gls{SDRi} after detection is slightly better but remains lower than the baseline.
We can also evaluate the \gls{AS}/\gls{IS} detection capability of the \gls{TSEIS} by looking at the detection performance of a classifier based on the attenuation as introduced in Eq.~\eqref{eq:classifier_att}. Figure \ref{fig:DET} shows the \gls{DET} curve and \gls{EER} of such a classifier.

The \gls{TSE} module in \gls{TSEV} corresponds to the above baseline system, and thus the performance before detection is the same for the \gls{AS} cases. The proposed verification based on the cosine distance of the embeddings computed with the auxiliary \gls{NN} is simple yet effective. Indeed, it can detect relatively well \gls{AS}/\gls{IS}, with an \gls{EER} of less than 9~\%. The \gls{SDRi} after detection is 0.5~dB lower because it includes miss detection errors.
The total error rate on \gls{AS} cases, i.e., Fail\&Miss rate, is 10.5~\%, which is better than the \gls{TSEIS} system by about 3~\%. Overall \gls{TSEV} achieves higher extraction and detection performance than \gls{TSEIS}.

We also explore the impact of training with a larger training set which includes more speakers with the \gls{TSEV}(360) system. Retraining on the larger training set improves \gls{SDRi} by about 1.2~dB, but mainly it can greatly reduce the failure rate, the \gls{EER} and the combined Fail\&Miss.

Figure \ref{fig:DET} plots the \gls{DET} curves for the \gls{AS}/\gls{IS} detection with \gls{TSEV} and \gls{TSEIS}. We observe that the miss rate rapidly increases for the \gls{TSEIS}, while the curve for  \gls{TSEV} is much smoother. Consequently, it is more challenging to tune  at test time the false alarm or miss rate of  \gls{TSEIS} than \gls{TSEV}.

\begin{figure}[tb]
\centering
  \centerline{\includegraphics[width=0.95\linewidth]{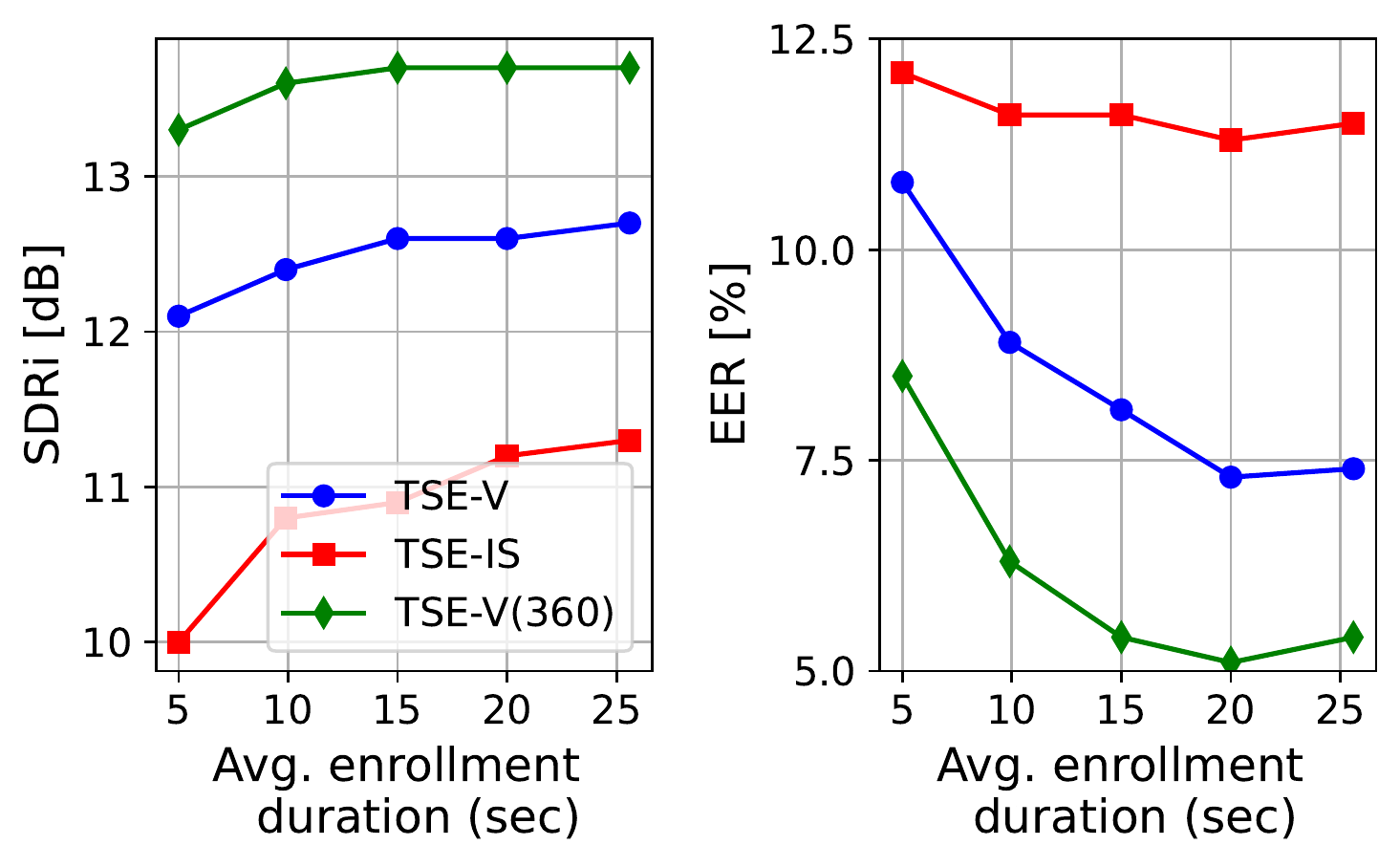}}
    \vspace{-3mm}
\caption{Extraction and \gls{AS}/\gls{IS} detection performance as a function of the enrollment duration.}
\vspace{-3mm}
\label{fig:enroll_duration}
\end{figure}

Finally, Figure \ref{fig:enroll_duration} plots the \gls{SDRi} before detection and \gls{EER} as a function of the average enrollment duration. We varied the enrollment length by concatenating from 1 to 5 enrollment utterances for each test sample, resulting in the average utterance length varying from 5 to 25 seconds. Longer enrollments improve extraction performance moderately but greatly reduce \gls{EER} for \gls{TSEV} and \gls{TSEV}(360). With  \gls{TSEV}(360), we can approach an \gls{EER} of 5~\% when using an enrollment utterance of 15 to 20 sec. 
\gls{TSEIS} does not exhibit a similar trend.

The results of our experiments demonstrate that with slight modifications, a \gls{TSE} system can handle \gls{IS} cases relatively well. 
The \gls{TSEV} approach provides better overall extraction and \gls{AS}/\gls{IS} detection performance than \gls{TSEIS}. It also allows more control to tune the miss detection and false alarm rates at test time. However, \gls{TSEV} requires an additional verification step. 
\gls{TSEIS} can learn to detect internally \gls{IS} cases and output directly zeros signals without increasing the computational complexity. Although \gls{TSEIS} performs worse than \gls{TSEV}, it could still be advantageous for, e.g., low-latency systems where the batch verification step would not be allowed. 

\section{Conclusion}
A \gls{TSE} system must perform speech extraction and speaker identification.
Most studies have focused on evaluating \gls{TSE}  systems in terms of extraction performance and have mostly ignored the impact of false alarms when the target speaker is inactive.
In this paper, we systematically compared two possible schemes to handle \gls{IS}.
Our experiments revealed that we could exploit the auxiliary \gls{NN} of a \gls{TSE} system to perform 
speaker verification at the output and detect \gls{AS}/\gls{IS} cases. We can detect \gls{AS}/\gls{IS} cases with a \gls{EER} of around 5~\%, using \gls{TSEV} trained with a relatively large amount of speaker, and using enrollment utterances of more than 15 sec. This positive finding confirms the potential of \gls{TSE} systems.

Our \gls{TSEV} system outperforms a \gls{TSEIS} system that can internally detect \gls{IS} and output zero signals. 
However, the \gls{TSEIS} system may remain attractive for, e.g., low-latency systems, which we plan to explore in our future works.

\bibliographystyle{IEEEtran}

\bibliography{mybib}
\end{document}